\documentclass[12pt]{article}
 
\usepackage[margin=1in]{geometry} 
\usepackage{amsmath,amsthm,amssymb}
\usepackage{graphicx}
\usepackage{float}
\usepackage{multirow}

\newcommand{\indep}{\rotatebox[origin=c]{90}{$\models$}}
\DeclareSymbolFont{symbolsC}{U}{pxsyc}{m}{n}
\DeclareMathSymbol{\coloneqq}{\mathrel}{symbolsC}{"42}

\newtheorem{assumption}{Assumption}
 
\begin{document}
 
\title{The LOOP Estimator: Adjusting for Covariates in Randomized Experiments}
\author{
	Edward Wu\thanks{Department of Statistics, University of Michigan,  Ann Arbor, MI.}
	\and
	Johann Gagnon-Bartsch\footnotemark[1]
}

\maketitle

\begin{abstract}
When conducting a randomized controlled trial, it is common to specify in advance the statistical analyses that will be used to analyze the data.  Typically these analyses will involve adjusting for small imbalances in baseline covariates.  However, this poses a dilemma, since adjusting for too many covariates can hurt precision more than it helps, and it is often unclear which covariates are predictive of outcome prior to conducting the experiment.  For example, both post-stratification and OLS regression adjustments can actually increase variance (relative to a simple difference in means) if too many covariates are used.  OLS is also biased under the Neyman-Rubin model.  In this paper, we introduce the LOOP (``Leave-One-Out Potential outcomes") estimator of the average treatment effect.  We leave out each observation and then impute that observation's treatment and control potential outcomes using a prediction algorithm, such as a random forest.  This estimator is unbiased under the Neyman-Rubin model, generally performs at least as well as the unadjusted estimator, and the experimental randomization largely justifies the statistical assumptions made.  Importantly, the LOOP estimator also enables us to take advantage of automatic variable selection when using random forests.

\end{abstract}

\section{Introduction}
In randomized controlled trials, it is common to specify in advance the statistical analyses to be performed. For example, various authors have advocated for the reporting of statistical methods in the trial protocol (\textit{e.g.}, \cite{Begg}, \cite{Schulz}). Typically these analyses will involve adjusting for small imbalances in baseline covariates, which can improve the precision of the treatment effect estimate.  However, in cases where the analysis methods are pre-specified, it can be unclear which covariates should be used and if covariate adjustment will even be helpful.  An overly aggressive adjustment that adjusts for too many covariates can hurt precision more than it helps. 
\\\\
Covariate adjustment is commonly done through regression; for example, Young \cite{Young} cites 53 experimental papers from the economics literature between 2007 and 2014 in which regression adjustment is used. Although it is common, OLS does have disadvantages. One of the virtues of randomized experiments is that the physical act of randomization largely justifies the statistical assumptions of the Neyman-Rubin model,\footnote{One important assumption that is not guaranteed by randomization is that one unit's outcome is not affected by another unit's treatment status.  This assumption is sometimes referred to as the Stable Unit Treatment Value Assumption (SUTVA).} a non-parametric model which was first introduced by Jerzy Neyman \cite{Neyman} and further developed by Donald Rubin \cite{Rubin}.  However, the Neyman-Rubin model is quite different than the standard OLS model, and as noted by Freedman \cite{Freedman}, randomization fails to justify the standard assumptions of OLS.  Moreover, Freedman shows that the regression estimate is biased under the Neyman-Rubin model and can, in certain circumstances, be outperformed by a simple difference in means.   In a response to Freedman's paper, Lin \cite{Lin} argues that ``without taking the regression model literally, we can still make use of properties of OLS that do not depend on the model assumptions." In other words, even when the regression model is incorrect, regression adjustment can be a useful tool. Both Freedman and Lin note that the OLS estimator performs well in large sample sizes. However, in cases where we have only a moderate sample size and a relatively large number of covariates, variable selection may be required.
\\\\
While regression is a common method of covariate adjustment, there are others. For example, Bloniarz et al.\ \cite{Bloniarz} propose the use of lasso adjustments when the number of covariates is large, especially when the number of covariates exceeds the number of experimental units. Another covariate adjustment method is post-stratification \cite{Holt}. Post-stratification is an adjustment made by stratifying on a pretreatment variable, estimating the treatment effect within each stratum, and taking the weighted average over all strata. Miratrix, Sekhon, and Yu \cite{Miratrix} explore the properties of the post-stratified estimator under the Neyman-Rubin model. Rosenbaum \cite{Rosenbaum} also discusses covariate adjustment in the context of randomization inference. Rosenbaum uses the covariates to estimate the control outcome for each unit, calculates residuals from these estimates, and permutes the residuals to test hypothesized values of the treatment effect. He then inverts the hypothesis tests to yield confidence intervals. Rosenbaum notes that one can obtain the residuals using any fitting algorithm and cites robust linear regression, rank linear regression, or a smoother as examples (in addition to OLS). While Rosenbaum's method relies only on randomization as the basis for inference, it assumes a fixed treatment effect for each unit.
\\\\
Aronow and Middleton \cite{Aronow} introduce another estimator, which is related to the Horvitz-Thompson estimator \cite{Horvitz-Thompson}. This design-based estimator involves the estimation of a function of the covariates. So long as this function is independent of the treatment assignment, the resulting estimate will be unbiased.  We propose a special case of this estimator, the LOOP (``Leave-One-Out Potential outcomes") estimator. We leave out each observation and then impute that observation's treatment and control potential outcomes using a prediction algorithm, such as a random forest \cite{Breiman}.  Our work is similar to that of Wager, Du, Taylor, and Tibshirani \cite{Wager}, who also propose a set of estimators that build on the work of Aronow and Middleton, and use machine learning methods to impute potential outcomes.  Wager et al.\ assume that the experimental units are drawn from a superpopulation, and focus primarily on the population average treatment effect.
\\\\
In this paper, we analyze the LOOP estimator assuming that the potential outcomes and the covariates are fixed and that the only source of randomness is in the treatment assignment. We derive an estimate for the variance of the LOOP estimator. Aronow and Middleton also provide an estimate for the variance of their estimator, but assume that the function of the covariates is constant.   Note that our variance estimate also differs from that of Wager et al., as we work under a different model.
\\\\
We discuss the imputation of each unit's potential outcomes using various methods such as decision trees.  We show that using the LOOP estimator and imputing potential outcomes using a decision tree is equivalent to post-stratification.  Because random forests are typically an improvement over individual decision trees, our hope is that we can use the LOOP estimator with random forests to improve upon post-stratification.  Miratrix et al.\ note that post-stratification is nearly as efficient as blocking, and we therefore hope to obtain an estimate that works as well or better than if we had used a blocked design.  
\\\\
To summarize, the primary advantages of the LOOP estimator are: (1) it is design-based, meaning that the experimental randomization largely justifies the statistical assumptions; (2) it is exactly unbiased; (3) it generally performs no worse than the simple difference-in-means estimator, but can often substantially improve performance; and importantly (4) it allows for automatic variable selection, so we do not need to know which covariates to use ahead of time. 
\\\\
The paper is organized as follows. Section \ref{motivation} provides a motivating example. In Section \ref{LOOP}, we introduce notation and assumptions and discuss the simple difference and LOOP estimators. In Section \ref{pomodels}, we discuss three different methods of imputing the potential outcomes and relate the LOOP estimator with imputation done by decision trees and random forests to post-stratification.  In Section \ref{dependent}, we discuss how to modify the procedures to account for different experimental designs such as block designs. In Section \ref{variance}, we provide an estimate of the variance. In Section \ref{results}, we apply the LOOP estimator to two examples: one using simulated data and one using real experimental data. Section \ref{discussion} concludes.

\section{Motivation} \label{motivation}
Our motivating example is a so-called ``pay for success'' program in the state of Illinois \cite{pfs}.  In brief, a pay for success program is one in which a government contracts an outside organization to provide needed services, but only pays the organization if the services are shown to be effective, typically in a randomized controlled experiment.  In our example, the contracted organization is to provide special social services to at-risk youth, and one metric for success (among others) is a reduction in the number of days spent in juvenile detention.  Success of the program will be evaluated according to the results of a six year experiment in which eligible youth are randomly selected to receive either the special services or ordinary care.  The evaluation will be conducted by researchers in the School of Social Work at the University of Michigan; author Gagnon-Bartsch of this paper assisted the evaluators in planning the design and analysis of the experiment.  Unfortunately, the experiment has only recently begun so we do not yet have any data on which to apply the methods we develop in this paper, and our discussion of the pay for success program is therefore limited to this section (we explore an alternative dataset in Section \ref{results}).  Nonetheless, the challenges presented by the pay for success program are instructive, and we outline them briefly.   
\\\\
Several hundred youth are expected to take part in the program.  Eligible participants are independently randomized to treatment or control, each with probability 1/2.  More elaborate designs were considered, but were too logistically challenging.  A key difficulty is the fact that the participants enter into the experiment continually over time, making designs such as blocking infeasible.  
\\\\
Several baseline covariates will be available, at least some of which (\textit{e.g.}, age) are known to be highly predictive of outcome.  It was agreed that some form of adjustment for these covariates was desirable, but initially there was no clear consensus on which adjustment procedure should be used or which covariates should be included.  Given the need to specify the analysis protocol in advance, this led to considerable discussion.  In the end, it was agreed to use a post-stratification estimator, partly on the grounds that it is unbiased under the Neyman-Rubin model, whereas other common estimators (\textit{e.g.}, linear regression) are not.  Unbiasedness is arguably more inherently desirable in this example than in many other applications because the state's payment rate for the services provided will be directly proportional to the estimated size of the treatment effect. Any bias in the estimator therefore effectively results in a bias in the payment.\footnote{Note that even if the estimator is unbiased, payment will still be biased for other reasons.  In particular, no payment will be made at all unless the observed treatment effect achieves statistical significance, which results in a payment bias against the service provider.  On the other hand, if the observed treatment effect turns out to be negative, the service provider will not receive a negative payment (\textit{i.e.}, will not be required to pay the state), which results in a bias in favor of the service provider.}  Moreover, it was agreed to post-stratify on just two variables that are known to be highly predictive of outcome; other covariates will not be used due to the risk that an overly aggressive adjustment could end up hurting precision rather than improving it.  
\\\\
In this paper, we are motivated to produce a method that provides automatic variable selection in order to eliminate the guesswork in deciding which covariates to use, while remaining unbiased under the Neyman-Rubin model.  An initial idea was to randomly split the data in half, use one half to empirically determine which covariates are predictive of outcome and construct a set of strata that are optimal in some sense (perhaps using a decision tree), and then use the other half of the data to compute a post-stratified estimate using the optimal strata.  Since the data used to construct the strata would be independent of the data used in the estimation step, the estimator would remain unbiased.  Only half of the data would be used in the estimation step, however, this procedure could then be repeated many times and the results averaged to produce an aggregate estimator that is also unbiased but effectively makes use of all of the data.  
\\\\
The method we develop in this paper, which is a special case of the estimator proposed by Aronow and Middleton \cite{Aronow}, is very similar in spirit the procedure just described.  It is in some sense a limiting case in which the data is split, not in half, but rather such that all of the observations except for one are used to determine the optimal strata, and (counterintuitively) only one observation is left for estimation.  Moreover, instead of relying on just one single set of optimal strata, we use many nearly optimal sets; these sets of strata are effectively determined by a random forest algorithm.

\section{The LOOP Estimator} \label{LOOP}
In this section, we introduce the LOOP (``Leave-One-Out Potential outcomes") estimator, which we can use to obtain an unbiased estimate of the average treatment effect while adjusting for covariates.

\subsection{Model and Notation}
Consider a randomized controlled experiment in which there are $N$ participants, indexed by $i = 1, 2, ..., N$.  Each participant is randomly assigned to either treatment or control, and we let $T_i$ denote the $i$-th participant's treatment assignment, such that $T_i = 1$ if the $i$-th participant is assigned to treatment and $T_i = 0$ if the $i$-th participant is assigned to control.  For each participant, we observe (in addition to the treatment assignment $T_i$) a response variable $Y_i$ and a $q$-dimensional vector of baseline covariates $Z_i$.
\\\\
We assume Bernoulli treatment assignments, \textit{i.e.},
\begin{equation}
T_i \indep T_j
\end{equation}
for $i \ne j$.  We let $p_i$ denote the $i$-th participant's probability of being assigned to treatment, \textit{i.e.},
\begin{equation}
p_i = P(T_i = 1)
\end{equation}
and assume $0 < p_i < 1$.  In some parts of this paper, we assume for simplicity (and without much loss of generality) that $p_i = p$ for all $i$ and for some fixed constant $p$, but for now we explicitly let $p_i$ vary from subject to subject.   
\\\\
Associated with each of the $N$ participants are two fixed (non-random) potential outcomes, $t_i$ and $c_i$.  We assume that we observe $t_i$ if participant $i$ is assigned to treatment and $c_i$ if participant $i$ is assigned to control. That is, the observed outcome $Y_i$ for participant $i$ is  
\begin{equation}
Y_i = T_it_i + (1-T_i)c_i.
\end{equation} 
We define the individual treatment effect $\tau_i$ as
\begin{equation}
\tau_i = t_i - c_i
\end{equation}
and the average treatment effect $\bar\tau$ as
\begin{equation}
\bar{\tau} = \frac{1}{N}\sum_{i=1}^{N} \tau_i
\end{equation}
which is our primary parameter of interest.
\\\\
Lastly, some additional notation.  Let $\mathcal{T} = \left\{i:  T_i = 1\right\} $ and $\mathcal{C} = \left\{i: T_i = 0\right\}$. Let $n$ be the (random) number of participants assigned to treatment and $N - n$ be the number assigned to control.  For each participant, we define the important quantity $m_i$ as 
\begin{equation}
m_i = (1-p_i) t_i + p_i c_i.
\end{equation}  
Note that when $p_i = \frac{1}{2}$, this is simply the mean of $t_i$ and $c_i$.  We will use the notation $\hat{m}_i$ to denote an estimate of $m_i$.  
Finally, we define the (signed) inverse probability weights $U_i$ as 
\begin{equation}
U_i = \left\{\begin{array}{lr} 1/{p_i}, & T_i = 1 \\ {-1}/{(1-p_i)}, & T_i = 0  \end{array}\right.
\end{equation}
and note that $U_i$ has expectation 0.

\subsection{Average and Individual Treatment Effects}
It is not possible to observe any single participant's treatment effect $\tau_i$, because for each participant we are only able to observe the treatment response $t_i$ or the control response $c_i$.  However, it is well known that the average treatment effect $\bar{\tau}$ can be estimated.  We define the \textit{simple difference estimator} $\hat{\tau}_{sd}$ to be the difference of the average of the observed treatment responses and the average of the observed control responses:
\begin{equation}
\hat{\tau}_{sd} = \frac{1}{n}\sum_{i \in \mathcal{T}}Y_i-\frac{1}{N-n}\sum_{i \in \mathcal{C}}Y_i.
\end{equation}
This provides an unbiased estimate of the average treatment effect (conditional on $0 < n < N$). 
\\\\
Less well known is the fact that it is also possible to provide an unbiased estimate of an individual participant's treatment effect $\tau_i$.  For example, $Y_iU_i$ is one such estimator:
\begin{equation}
Y_iU_i = \left\{\begin{array}{lr} t_i/{p_i}, & T_i = 1 \\ {-c_i}/{(1-p_i)}, & T_i = 0  \end{array}\right.
\end{equation}
and thus 
\begin{align}
\mathbb{E}(Y_iU_i) &= \frac{t_i}{p_i}P(T_i=1) + \frac{-c_i}{1-p_i}P(T_i=0) \\
&= t_i - c_i.
\end{align}
This estimator is essentially mathematical trickery.  Suppose, for example, that $p_i = 1/2$.  Then if participant $i$ is assigned to treatment we would estimate his treatment effect as $2Y_i$, and if he was assigned to control we would estimate his treatment effect as $-2Y_i$.  Although this does result in an unbiased estimator of $\tau_i$, it is clearly useless for all practical purposes.  A more sanguine way of putting this would be that the estimator, despite being unbiased, likely has very high variance.
\\\\
As an alternative estimator of $\tau_i$, consider 
\begin{equation}
\hat{\tau}_i = (Y_i - \hat{m}_i)U_i.  \label{tauidef}
\end{equation}
If $\hat{m}_i$ is independent of $U_i$ --- that is, if $\hat{m}_i$ is independent of the $i$-th participant's treatment assignment --- then $\hat{\tau}_i$ is an unbiased estimator of $\tau_i$:
\begin{align} 
\mathbb{E}(\hat{\tau}_i) &= \mathbb{E}\left[(Y_i - \hat{m}_i)U_i\right] \nonumber \\ 
                         &= \mathbb{E}(Y_iU_i)  - \mathbb{E}(\hat{m}_i)\mathbb{E}(U_i) \nonumber \\
                         &= \tau_i
\end{align}
where in the last line we use the fact that $\mathbb{E}(U_i) = 0$.
The advantage of this estimator is that it will have a low variance as long as $\hat{m}_i \approx m_i$.  To see why, suppose that $\hat{m}_i = m_i$ exactly.  Then
\begin{equation}
(Y_i-m_i)U_i = \left\{\begin{array}{lr} (t_i-m_i)/{p_i}, & T_i = 1 \\ {(-c_i + m_i)}/{(1-p_i)}, & T_i = 0  \end{array}\right.
\end{equation}
but both $(t_i-m_i)/{p_i}$ and $(-c_i + m_i)/(1-p_i)$ work out to be $\tau_i$, and thus $\hat{\tau}_i$ is not only unbiased but also has zero variance.  When $\hat{m}_i$ only approximately equals $m_i$, then the variance of $\hat{\tau}_i$ is no longer zero but is small.  More precisely, in Section \ref{variance} we show that 
\begin{equation}
\mathrm{Var}(\hat{\tau}_i) = \frac{1}{p_i(1-p_i)}\mathbb{E}\left[(\hat{m}_i - m_i)^2\right].
\end{equation}
To summarize then, $\hat{\tau}_i$ will be unbiased and have low variance as long as: (1) $\hat{m}_i$ is independent of $T_i$; and (2) $\hat{m}_i$ is a good estimator of $m_i$.

\subsection{Leave-One-Out Imputation}

We now define the LOOP estimator of the average treatment effect $\bar{\tau}$ as: 
\begin{equation}
\hat{\tau} = \frac{1}{N}\sum_{i=1}^{N}\hat{\tau}_i
\end{equation}
where $\hat{\tau}_i$ is defined as in (\ref{tauidef}) and where $\hat{m}_i$ is obtained as follows.  For each $i$, we drop observation $i$ and use the remaining $N-1$ observations to impute $t_i$ and $c_i$, using any method of our choosing (\textit{e.g.}, linear regression, random forests, etc.).  Having obtained estimates $\hat{t}_i$ and $\hat{c}_i$, we then set 
\begin{equation}
\hat{m}_i = (1-p_i) \hat{t}_i + p_i\hat{c_i}. \label{mhati}
\end{equation}
As an example, suppose we wish to estimate $\hat{m}_i$ using linear regression.  For each $i$, we would drop observation $i$ and then regress $Y$ on $T$ and $Z$ using only the remaining $N-1$ observations.  We would then calculate $\hat{t}_i$ and $\hat{c}_i$ using the fitted model, plugging in $Z_i$ for the covariates, and then compute $\hat{m}_i$ as in (\ref{mhati}).
\\\\
Because we leave out the $i$-th observation when we compute $\hat{m}_i$, it follows that $T_i$ and $\hat{m}_i$ are independent and thus that $\hat{\tau}_i$ is unbiased.  It immediately follows that $\hat{\tau}$ is also unbiased.  This will be true no matter how we estimate $t_i$ and $c_i$, as long as we leave out observation $i$ so that $\hat{t}_i$ and $\hat{c}_i$ are independent of $T_i$.  Importantly, note that we impute both $t_i$ and $c_i$, even though one of them is actually observed and therefore known.  If we were to use the true observed value, then $\hat{m}_i$ would no longer be independent of $T_i$.
\\\\
It is worth noting that although we use the individual treatment effect estimates $\hat{\tau}_i$ in this paper simply as an intermediate step in the estimation of the average treatment effect $\bar{\tau}$, these individual treatment effect estimates may be useful for other purposes as well, such as in estimating treatment effect heterogeneity.  With this in mind, we summarize below three useful facts about the $\hat{\tau}_i$, the latter two of which we show in Section \ref{variance}: 
\begin{align}
\mathbb{E}(\hat{\tau}_i) &= \tau_i \\
\mathrm{Var}(\hat{\tau}_i)  &= \frac{1}{p_i(1-p_i)}\mathbb{E}\left[(\hat{m}_i - m_i)^2\right] \\
\mathrm{Cov}(\hat{\tau}_i, \hat{\tau}_j) &= \mathrm{Cov}(\hat{m}_iU_i,\hat{m}_jU_j).
\end{align}
The covariance term $\mathrm{Cov}(\hat{m}_iU_i,\hat{m}_jU_j)$ is usually negligible and can be ignored in most applications (note that $U_i$ and $U_j$ are independent).

\section{Imputing the Potential Outcomes} \label{pomodels}
In the subsequent sections, we propose several methods for imputing the potential outcomes in order to estimate $m_i$. First, we impute the potential outcomes without making use of covariates, simply taking the mean of the observed outcomes in each treatment group.  When we do this, we see that the LOOP estimator is exactly equal to the simple difference estimator.  We also impute the potential outcomes using decision trees and discuss the connection between post-stratification and the LOOP estimator. Finally, we propose the use of random forests, which may provide an improvement over post-stratification and allow us to take advantage of automatic variable selection.

\subsection{Imputing Potential Outcomes Ignoring Covariates:\\
LOOP equals the Simple Difference Estimator} \label{imputing1}
In this section, we impute the potential outcomes without making use of covariates.  We simply take the mean of the observed outcomes in the treatment group (excluding observation $i$) to estimate $t_i$ and the mean of the observed outcomes in the control group (excluding observation $i$) to estimate $c_i$.  If the assignment probabilities are all equal, \textit{i.e.}, if $p_i = p$ for all $i$ and for some fixed $p$, then the LOOP estimator is exactly equivalent to the simple difference estimator, as we show below:
	\begin{align} 
        \hat{\tau} & = \frac{1}{N}\sum_{i=1}^{N}\left(Y_i - \hat{m}_i\right)U_i \nonumber \\ 
        & = \frac{1}{N}\left[\sum_{i=1}^{N}\frac{1}{p}\left(Y_i - \hat{m}_i\right)T_i +  \sum_{i=1}^{N}\frac{1}{1-p}\left(\hat{m}_i - Y_i\right)(1-T_i)\right] \nonumber \\ 
	& = \frac{1}{N}\left\{\sum_{i=1}^{N}\frac{1}{p}\left[Y_i - \left(\frac{\sum_{k \in \mathcal{T}\backslash{}i}(1-p)Y_k}{n-T_i}+\frac{\sum_{k \in \mathcal{C}\backslash{}i}pY_k}{(N-n)-(1-T_i)}\right)\right]T_i + \right. \nonumber
	\\ 
	& \left.\hspace*{1.5cm} \sum_{i=1}^{N}\frac{1}{1-p}\left[\left(\frac{\sum_{k \in \mathcal{T}\backslash{}i}(1-p)Y_k}{n-T_i}+\frac{\sum_{k \in \mathcal{C}\backslash{}i}pY_k}{(N-n)-(1-T_i)}\right) - Y_i\right](1-T_i)\right\} \nonumber \\ 
	& = \frac{1}{N}\left[\sum_{i \in \mathcal{T}}\left(\frac{Y_i}{p} - \frac{1-p}{p}\frac{\sum_{k \in \mathcal{T}\backslash{}i}Y_k}{n-1}-\frac{\sum_{k \in \mathcal{C}}Y_k}{N-n}\right) + \sum_{i \in \mathcal{C}}\left(\frac{\sum_{k \in \mathcal{T}}Y_k}{n}+\frac{p}{1-p}\frac{\sum_{k \in \mathcal{C}\backslash i}Y_k}{(N-n)-1} - \frac{Y_i}{1-p}\right)\right] \nonumber \\
	& = \frac{1}{N}\left[\sum_{i \in \mathcal{T}}\frac{Y_i}{p} - \sum_{i \in \mathcal{C}}\frac{Y_i}{1-p}-\frac{1-p}{p}\frac{(n-1)\sum_{k \in \mathcal{T}}Y_k}{n-1}-\frac{n\sum_{k \in \mathcal{C}}Y_k}{N-n} + \frac{(N-n)\sum_{k \in \mathcal{T}}Y_k}{n} \right. \nonumber \\ 
	& \left. \hspace*{1.5cm} +\frac{p}{1-p}\frac{((N-n)-1)\sum_{k \in \mathcal{C}}Y_k}{(N-n)-1}\right] \nonumber \\
	& = \frac{1}{N}\left[\sum_{i \in \mathcal{T}}\frac{Y_i-(1-p)Y_i}{p} - \sum_{i \in \mathcal{C}}\frac{Y_i-pY_i}{1-p}-\frac{n\sum_{k \in \mathcal{C}}Y_k}{N-n} + \frac{(N-n)\sum_{k \in \mathcal{T}}Y_k}{n}\right] \nonumber \\ 
	& = \frac{1}{N}\left[\sum_{i \in \mathcal{T}}Y_i - \sum_{i \in \mathcal{C}}Y_i-\frac{n\sum_{k \in \mathcal{C}}Y_k}{N-n} + \frac{(N-n)\sum_{k \in \mathcal{T}}Y_k}{n}\right] \nonumber \\ 
	& = \frac{1}{N}\left[\frac{((N-n)+n)\sum_{k \in \mathcal{T}}Y_k}{n}-\frac{(n+(N-n))\sum_{k \in \mathcal{C}}Y_k}{N-n}\right] \nonumber
	\\ &
	= \frac{\sum_{k \in \mathcal{T}}Y_k}{n}-\frac{\sum_{k \in \mathcal{C}}Y_k}{N-n} \nonumber \\
	& =\hat{\tau}_{sd}.
	\end{align}
As a result of this equivalence, we conclude that in practice the LOOP estimator will typically perform no worse than the simple difference estimator.  That is, the LOOP estimator will outperform the simple difference estimator as long as we improve the imputation of the potential outcomes beyond this baseline approach.  In particular, we find it reassuring that the leave-one-out procedure does not inherently introduce extra variance.  
\\\\
Technical note: One minor difference between the simple difference estimator and the LOOP estimator in this case is that the simple difference estimator is undefined whenever $n$ is equal to 0 or $N$, whereas the LOOP estimator is undefined whenever $n$ is equal to 0, 1, $N-1$, or $N$.

\subsection{Imputing Potential Outcomes using Decision Trees:\\
LOOP equals Post-stratification}
In this section, we discuss the connection between the LOOP estimator and post-stratification. Post-stratification is a covariate adjustment method made by stratifying on pretreatment variables, estimating the treatment effect within each stratum by taking a simple difference in means, and then taking the weighted average over all strata \cite{Miratrix}. We argue that when we impute potential outcomes using a decision tree, the LOOP estimator is equivalent to post-stratification. 
\\\\
Given a single decision tree (fixed in advance), we impute the potential outcomes as follows.  First, we assign each observation $i$ to a group; this is done by applying the decision tree to observation $i$'s covariates.  (This group may be viewed as a ``leaf'' or a ``stratum.'')  For each $i$, we then impute $t_i$ using the average observed outcome of the treated units within the same group (excluding observation $i$ itself).  We impute $c_i$ similarly.  Thus, using the same argument given above in Section \ref{imputing1}, it is simple to show that the average of the $\hat{\tau}_i$ within a group is equal to the simple difference within that group. Thus, the average of all the $\tau_i$ is a weighted average of the within-group simple differences, \textit{i.e.}, it is a post-stratification estimator.

\subsection{Imputing Potential Outcomes using Random Forests}

In their analysis of post-stratification, Miratrix et al.\ show that post-stratification is nearly as efficient as blocking. However, one disadvantage of post-stratification is that we must be parsimonious in the number of variables selected. If we include too many covariates, we end up partitioning our data too finely. We can overcome this limitation and also improve on the post-stratified estimate using the LOOP estimator. One advantage of the LOOP estimator is that estimation of $m_i$ is very flexible. One can impute the potential outcomes using any method, so long as $\hat{m}_i$ and $T_i$ are independent.  In particular, we can use ensemble methods such as boosting or bagging to improve our estimates over a single decision tree. 
\\\\
One such method is the random forest algorithm, and random forests will be our method of choice for imputing the potential outcomes for the remainder of the paper. In order to impute the potential outcomes using random forests, we could first omit observation $i$, and then create a random forest using the remaining $N-1$ observations, which we could use to impute $t_i$ and $c_i$.  However, this would be computationally demanding.  Fortunately, it is also unnecessary.  Random forests are naturally suited for the LOOP estimator. Although we describe a leave-one-out procedure, we can make use of the out-of-bag predictions in practice.  We can therefore fit a single random forest. For each $i$, we predict $c_i$ and $t_i$ using the out-of-bag predictions, \textit{i.e.,} using only the trees that do not include observation $i$. By contrast, when imputing the potential outcomes using many other methods, such as OLS, we do need to create a separate model for each $i$. As a result, imputing the potential outcomes with random forests can be relatively computationally efficient.
\\\\
Because random forests are typically an improvement over individual decision trees, they allow us to obtain a more precise estimate of the ATE. By using random forests to effectively improve upon post-stratification, we might even hope to obtain an estimate of the ATE that works as well as or better than if we had used a blocked experimental design.  Moreover, random forests essentially provide automatic variable selection, making it unnecessary to decide in advance which covariates should be used. Biau \cite{Biau} shows that the rate of convergence of the random forest algorithm depends on the number of important variables present, rather than how many noise variables there are.

\section{Variance Estimation} \label{variance}
Aronow and Middleton \cite{Aronow} give a conservative estimate of the variance of the Horvitz-Thompson estimator. They also provide an estimate for the variance of their own estimator, but only when the function of the covariates (\textit{i.e.}, our $\hat{m}_i$) is a constant fixed in advance, not computed from the data.  In this section, we derive an estimate for the variance of the LOOP estimator. Given the leave-one-out method we use to impute potential outcomes, the jackknife would be an obvious choice for estimating the variance. As Efron and Stein show, the jackknife variance estimate tends to be conservative \cite{Efron}. However, we found this estimate to be excessively conservative in the presence of treatment effect heterogeneity. Here we provide a different estimate for the variance of our estimator.  In Section \ref{truevariance} we calculate the true variance of $\hat{\tau}$ and then in Section \ref{estimatedvariance} we produce an estimate.
\subsection{Variance of $\hat{\tau}$} \label{truevariance}
We will show that 
\begin{align}
\mathrm{Var}(\hat{\tau}_i) = \frac{1}{p_i(1-p_i)}\mathrm{MSE}(\hat{m_i}) \label{vartauihat}
\end{align}
and that 
\begin{align}
	\gamma_{ij} = \mathrm{Cov}(\hat{\tau}_i,\hat{\tau}_j) = \rho_{ij}  \sqrt{\frac{\mathrm{Var}(\hat{m}_i)\mathrm{Var}(\hat{m}_j)}{p_i p_j (1-p_i)(1-p_j)}}  \label{gammaijexpression}
\end{align}
where
\begin{align}
	\rho_{ij} &= \mathrm{Corr}(\hat{m}_iU_i, \hat{m}_jU_j).
\end{align}
From these, it follows that
\begin{equation}
	\mathrm{Var}(\hat{\tau}) = \frac{1}{N^2}\left[\sum_{i=1}^{N}\frac{1}{p_i(1-p_i)}\mathrm{MSE}(\hat{m}_i)
	+
	\sum_{i\ne j}\gamma_{ij}
	\right]. \label{varexpr}
\end{equation}
We begin with the variance of a single $\hat{\tau}_i$:
\begin{align} 
	\mathrm{Var}(\hat{\tau_i}) & = \mathrm{Var}\left[\mathbb{E}(\hat{\tau_i}|\hat{m_i})\right] + \mathbb{E}[\mathrm{Var}(\hat{\tau_i}|\hat{m_i})] \nonumber
	\\
	& = \mathrm{Var}(\tau_i) + \mathbb{E}\left[\mathrm{Var}\left(\frac{1}{p_i}(Y_i - \hat{m_i})T_i + \frac{1}{1-p_i}(\hat{m_i} - Y_i)(1-T_i)|\hat{m_i}\right)\right] \nonumber
	\\
	& = 0 + \mathbb{E}\left[\mathrm{Var}\left(\frac{1}{p_i}(t_i - \hat{m_i})T_i + \frac{1}{1-p_i}(\hat{m_i} - c_i)(1-T_i)|\hat{m_i}\right)\right] \nonumber
	\\
	& =\frac{1}{p_i^2(1-p_i)^2}\mathbb{E}\left[\mathrm{Var}((1-p_i)(t_i - \hat{m_i})T_i + p_i(\hat{m_i} - c_i)(1-T_i)|\hat{m_i})\right] \nonumber
	\\
	& =\frac{1}{p_i^2(1-p_i)^2}\mathbb{E}[\mathrm{Var}[((1-p_i)t_i + p_ic_i - \hat{m_i})T_i + p_i(\hat{m_i} - c_i)|\hat{m_i}]] \nonumber
	\\
	& =\frac{1}{p_i^2(1-p_i)^2}\mathbb{E}[\mathrm{Var}[(m_i - \hat{m_i})T_i + p_i(\hat{m_i} - c_i)|\hat{m_i}]] \nonumber
	\\
	& =\frac{1}{p_i^2(1-p_i)^2}\mathbb{E}[(m_i - \hat{m_i})^2\mathrm{Var}(T_i|\hat{m_i})] \nonumber
	\\
	& = \frac{1}{p_i(1-p_i)}\mathbb{E}[(m_i - \hat{m_i})^2] \nonumber
	\\
	& = \frac{1}{p_i(1-p_i)}\mathrm{MSE}(\hat{m_i}). \label{vartau}
\end{align}
We now analyze the covariance term.  
\begin{align}
\gamma_{ij}
& = \mathrm{Cov}[(Y_i-\hat{m}_i)U_i,(Y_j-\hat{m}_j)U_j] \nonumber \\
& = \mathrm{Cov}(Y_iU_i,Y_jU_j) - \mathrm{Cov}(Y_iU_i,\hat{m}_jU_j) \nonumber \\
& \hspace{.5cm}
- \mathrm{Cov}(\hat{m}_iU_i, Y_jU_j) + \mathrm{Cov}(\hat{m}_iU_i, \hat{m}_jU_j).
\end{align}
The first term is zero, as $Y_iU_i$ and $Y_jU_j$ are independent. The second and third terms are also zero; for example, in the case of the second term,  
\begin{align}
	\mathrm{Cov}(Y_iU_i,\hat{m}_jU_j) & = \mathbb{E}(Y_iU_i\hat{m}_jU_j) - \mathbb{E}(Y_iU_i)\mathbb{E}(\hat{m}_jU_j) \nonumber \\
	& = \mathbb{E}(Y_iU_i\hat{m}_j)\mathbb{E}(U_j) - \mathbb{E}(Y_iU_i)\mathbb{E}(\hat{m}_j)\mathbb{E}(U_j) \nonumber \\
	& = 0
\end{align}
and a similar argument applies to the third term.
Thus,
\begin{align}
\gamma_{ij} & = \mathrm{Cov}(\hat{m}_iU_i, \hat{m}_jU_j) \nonumber \\
& = \rho_{ij} \sqrt{\mathrm{Var}(\hat{m}_iU_i)\mathrm{Var}(\hat{m}_jU_j)}  \nonumber \\
& = \rho_{ij} \sqrt{\frac{\mathrm{Var}(\hat{m}_i)\mathrm{Var}(\hat{m}_j)}{p_i p_j (1-p_i)(1-p_j)}}. \label{covtau}
\end{align}
Combining the results from (\ref{vartau}) and (\ref{covtau}) yield (\ref{varexpr}). Limiting our attention to the special case that $p_i = p$ for all $i$, 
\begin{equation}
\mathrm{Var}(\hat{\tau}) = \frac{1}{N}\left[\frac{\overline{\mathrm{MSE}}}{p(1-p)}  
+
(N-1)\bar{\gamma}
\right] \label{exacttauhatvar}
\end{equation}
where 
\begin{equation}
\overline{\mathrm{MSE}} = \frac{1}{N}\sum_{i=1}^{N}\mathrm{MSE}(\hat{m}_i)
\end{equation}
and
\begin{equation}
\bar{\gamma} = \frac{1}{N(N-1)}\sum_{i \ne j}\gamma_{ij}.
\end{equation}
In many cases, $\gamma_{ij}$ is negligible in the sense that $\gamma_{ij}$ (and likewise $\bar{\gamma}$) goes to zero faster than $1/N$, in which case
\begin{align}
\mathrm{Var}(\hat{\tau}) \approx \frac{\overline{\mathrm{MSE}}}{Np(1-p)}.
\end{align}
For example, suppose that under suitable regularity conditions $\mathrm{Var}(\hat{m}_i)$ and $\mathrm{Var}(\hat{m}_j)$ go to zero at rate $1/N$.  Then if $\rho_{ij}$ goes to zero (at any rate), $\gamma_{ij}$ will go to zero faster than $1/N$. Section \ref{gammabar} of the appendix gives a more formal argument: under 
conditions discussed in the appendix, we show that if $\bar{\rho} = \sum_{i\ne j}\rho_{ij}/[N(N-1)]$ goes to zero sufficiently quickly, $\bar{\gamma}$ will go to zero faster than $1/N$.
\\\\
To see why we might expect $\rho_{ij}$ (and likewise $\bar{\rho}$) to go to zero, recall that $U_i$ and $U_j$ are independent.  Thus, even if $\hat{m}_i$ and $\hat{m}_j$ are correlated (which they typically will be), $\rho_{ij}$ may still be negligible.  Indeed, 
if $\hat{m}_i$ and $\hat{m}_j$ are perfectly correlated, then $\rho_{ij} = 0$.  The only reason for $\hat{m}_iU_i$ and $\hat{m}_jU_j$ to be correlated would be through the dependence of $\hat{m}_i$ on $U_j$, and of $\hat{m}_j$ on $U_i$.  These dependencies will typically decay as $N$ grows.  As an illustrative example, suppose that for all $i$, $\hat{m}_i$ is a linear estimator, \textit{i.e.},\ for some constants $a_{i.k}$
\begin{equation}
\hat{m}_i = a_{i.0} + \sum_{k \ne i} a_{i.k} U_k.
\end{equation}
In this case, it can be shown (see Section \ref{polynomialcorrelation} of the appendix) that $\bar{\rho}$ goes to 0 at rate $1/N$; more specifically, we show $\bar{\rho} \le 1/(N-1)$.  Indeed, we further show (Section \ref{polynomialcorrelation}) that if $\hat{m}_i$ is a polynomial function of degree $D$ for all $i$, then $\bar{\rho} \le D/(N-1)$.
\\\\
Note that there do exist certain pathological cases where $\bar{\rho}$ can be large. For example, suppose that for all $i$, $\hat{m}_i = \prod_{k \neq i}U_k$. Then $\hat{m}_iU_i = \prod_{k=1}^{N}U_k$ for all $i$, so the correlation between $\hat{m}_iU_i$ and $\hat{m}_jU_j$ is exactly 1.  
\\\\
Fortunately, as we show in Section \ref{Var} of the appendix, $\gamma_{ij}$ (and thus $\bar{\gamma}$) is estimable, and we provide an explicit unbiased estimator.  Thus, if there is concern that in a particular application $\bar{\gamma}$ is not negligible --- either due to concern that $\bar{\gamma}$ may not go to zero faster than $1/N$ or simply due to concern that $N$ is not large enough --- it is not actually necessary to ignore the $\bar{\gamma}$ term when estimating the variance of $\hat{\tau}$.

\subsection{Estimating the Variance using Cross Validation}  \label{estimatedvariance}
In this section, we estimate the variance of the LOOP estimator using cross validation. Once again, we assume that $p_i = p$ for all $i$, so 
\begin{equation}\mathrm{Var}(\hat{\tau}) \approx \frac{1}{N}\left[\frac{1}{N}\sum_{i=1}^{N}\frac{1}{p(1-p)}\mathrm{MSE}(\hat{m}_i)\right].
\end{equation}
We can bound the MSE of $\hat{m}_i$ in terms of the MSEs of $\hat{t}_i$ and $\hat{c}_i$ (see Appendix \ref{bound} for the derivation):
\begin{align} 
	\mathrm{MSE}(\hat{m}_i) & \leq (1-p)^2\mathrm{MSE}(\hat{t_i})+p^2\mathrm{MSE}(\hat{c_i})+2p(1-p)\sqrt{\mathrm{MSE}(\hat{t_i})\mathrm{MSE}(\hat{c_i})}.  \label{msebound}
\end{align}
We can therefore bound the variance of $\hat{\tau}$ as follows:
\begin{align} 
	\mathrm{Var}(\hat{\tau}) & \approx\frac{1}{N}\left[\frac{1}{N}\sum_{i=1}^{N}\frac{1}{p(1-p)}\mathrm{MSE}(\hat{m}_i)\right] \nonumber 
	\\ & \leq \frac{1}{N}\left[\frac{1-p}{p}\frac{1}{N}\sum_{i=1}^{N}\mathrm{MSE}(\hat{t_i})+\frac{p}{1-p}\frac{1}{N}\sum_{i=1}^{N}\mathrm{MSE}(\hat{c_i})+2\frac{1}{N}\sum_{i=1}^{N}\sqrt{\mathrm{MSE}(\hat{t_i})\mathrm{MSE}(\hat{c_i})}\right] \nonumber
	\\ & \leq \frac{1}{N}\left[\frac{1-p}{p}\frac{1}{N}\sum_{i=1}^{N}\mathrm{MSE}(\hat{t_i})+\frac{p}{1-p}\frac{1}{N}\sum_{i=1}^{N}\mathrm{MSE}(\hat{c_i})+2\sqrt{\frac{1}{N}\sum_{i=1}^{N}\mathrm{MSE}(\hat{t_i})\frac{1}{N}\sum_{i=1}^{N}\mathrm{MSE}(\hat{c_i})}\right] \nonumber
    \\ & = \frac{1}{N}\left[\frac{1-p}{p}M_t+\frac{p}{1-p}M_c+2\sqrt{M_t M_c}\right] \label{varbound}
\end{align}
where 
\begin{equation}
M_t = \frac{1}{N}\sum_{i=1}^{N}\mathrm{MSE}(\hat{t_i})
\end{equation}
and
\begin{equation}
M_c = \frac{1}{N}\sum_{i=1}^{N}\mathrm{MSE}(\hat{c_i}).
\end{equation}
We estimate $M_t$ and $M_c$ by leave-one-out cross validation: 
\begin{align} 
\hat{M}_t = & \frac{1}{n}\sum_{i \in \mathcal{T}}(\hat{t}_i - t_i)^2 \label{Mhatt}
\\
\hat{M}_c = & \frac{1}{N-n}\sum_{i \in \mathcal{C}}(\hat{c}_i - c_i)^2. \label{Mhatc}
\end{align}
In Appendix \ref{MSEestimation}, we show that these estimates are nearly unbiased.  (Changing the denominator in (\ref{Mhatt}) from $n$ to $Np$ and the denominator in (\ref{Mhatc}) from $N-n$ to $N(1-p)$ results in estimators that are exactly unbiased.) We plug (\ref{Mhatt}) and (\ref{Mhatc}) into (\ref{varbound}) to obtain our final variance estimate:
\begin{align} 
	\widehat{\mathrm{Var}}(\hat{\tau}) =  \frac{1}{N}\left[\frac{1-p}{p}\hat{M}_t + \frac{p}{1-p}\hat{M}_c + 2\sqrt{\hat{M}_t\hat{M}_c}\right].
\end{align}
Finally, we note that when we impute potential outcomes ignoring covariates (as in Section \ref{imputing1}), $\hat{M}_t$ is simply the standard sample variance (of the treated units) times $\frac{n}{n-1}$ (and similarly for $\hat{M}_c$). That is, as we show in Appendix \ref{M_t Sample Var}, if we set $\hat{t}_i = \frac{1}{n - T_i}\sum_{j \in \mathcal{T}\backslash i}Y_j$, then $\hat{M}_t = \frac{n}{(n-1)^2}\sum_{i \in \mathcal{T}}(t_i - \bar{t})^2$.

\section{Dependent Treatment Assignments} \label{dependent}
In the preceding sections, we assumed that the treatment assignments are independent of each other. It is common for researchers to randomly assign a fixed number $n$ of participants to treatment and leave the remaining $N-n$ as controls. In such cases, treatment assignments are not independent. However, we can ensure the independence of $T_i$ and $\hat{m}_i$ as follows: if the $i$-th observation is assigned to treatment, we randomly pick one of the control observations and drop that observation as well as observation $i$ when fitting our imputation model.  Conversely, if the $i$-th observation is control, we randomly drop one of the treatment observations.
Thus, regardless of whether $T_i$ is equal to 0 or 1, when we estimate $\hat{m}_i$, we use $N-2$ of the remaining $N-1$ observations.  Of these $N-2$ observations, $n - 1$ will be assigned to treatment, $N-n-1$ will be assigned to control, and the specific allocation will be independent of $T_i$. We give an example to illustrate this ``random drop" procedure in Appendix \ref{RandomDropEx}. 
\\\\
Because this procedure ensures that $\hat{m}_i$ and $T_i$ are independent, $\hat{\tau}_i$ will remain unbiased.\footnote{In addition, the independence of $\hat{m}_i$ and $T_i$ also implies that (\ref{vartauihat}) continues to hold.  However, (\ref{gammaijexpression}) is no longer valid, due to the dependence of $U_i$ and $U_j$.  Variance estimation in this context may therefore require a modified approach.} By dropping an extra observation we are losing some information.  However, we could repeat this entire procedure many times, producing an unbiased estimate of $\hat{\tau}_i$ each time, which we could then average.  In the aggregate, we would then make use of all remaining $N-1$ observations. Note that in practice, the use of the random drop procedure would not change our estimates much. For example, if we use the random drop procedure with a decision tree, we would still obtain the post-stratified estimate. (See Appendix \ref{RandomDrop} for further discussion.)
\\\\
Note that a similar procedure could be used in a block-randomized experiment, in which a fixed number of participants within each block are assigned to treatment, and the rest to control.  In this case, when computing $\hat{m}_i$, we would need to drop an observation that is in the same block as $i$.  This procedure could even be extended to paired designs.  In a paired design, both observation $i$ and observation $i$'s pair would need to be dropped.  However, all of the remaining observations from the experiment could still be used to produce an estimate of $m_i$.

\section{Results} \label{results}
Below, we apply the LOOP estimator (with random forests) to both simulated and actual data. In our first simulation, we provide an illustrative example to demonstrate the bias of the point estimate and standard error for the OLS estimator. We also consider a simulation in which the response is binary. Finally, we apply the LOOP estimator to the experiment conducted by Barrera-Osorio et al.\ \cite{Barrera} on the effects of various cash transfer programs on educational outcomes in Colombia.

\subsection{Simulation 1: The OLS Estimate is Biased}
Consider a randomized experiment in which there are $N = 30$ subjects and there is a single covariate, $Z$, with three possible values: 0, 1, and 2. For each value of $Z$, there are 10 subjects and each subject has potential outcomes that are generated from a normal distribution with standard deviation 0.1. For $Z = 0$, the control and treatment outcomes have expectations 0 and 1, respectively; for $Z = 1$, the control and treatment outcomes both have expectation 1; and for $Z = 2$, the control and treatment outcomes have expectations 1 and 2. 
\\\\
After generating the treatment and control potential outcomes for the 30 subjects (which we do only once), we create 100,000 random assignment vectors ($T$) and the 100,000 corresponding vectors of observed outcomes ($Y$).  For each of these, we estimate the average treatment effect and nominal standard error. Below, we compare the results using OLS, the LOOP estimator with random forests, and cross estimation \cite{Wager} with random forests.\footnote{We use the code provided by \cite{Wager}; however, we remove the specified node size parameter.  This modification improves performance in the context of this simulation.} The bias is estimated as the mean point estimate minus the true ATE. We also show the mean nominal standard error and estimate the true standard error using the standard deviation of the 100,000 point estimates. The nominal standard errors for the LOOP estimator are calculated using the method of Section \ref{variance}, while the nominal standard errors for cross estimation are calculated using the estimator provided by \cite{Wager}. For OLS, the nominal standard errors are calculated using the usual formulas. 

\begin{table}[!htbp] \centering 
	\caption{Simulation Results: LOOP, Cross Estimation, and OLS} 
	\label{} 
	\begin{tabular}{@{\extracolsep{5pt}}lccc} 
		\\[-1.8ex]\hline 
		\hline \\[-1.8ex] 
	\textbf{Method} & \centering\textbf{Bias Estimate} & \centering\textbf{Mean Nominal SE} & \textbf{Estimate of True SE}\\ 
	\multirow{1}{*}{LOOP - RF} & \centering0.00006 &  \centering0.0442 & 0.0384\\ 
	\multirow{1}{*}{Cross Estimation - RF} & \centering 0.00067 & \centering0.1060 & 0.0373\\ 
	\multirow{1}{*}{OLS} & \centering-0.01415 & \centering0.1076 & 0.0440\\ \hline \\[-1.8ex] 
	\end{tabular} 
	\small Note: The bias estimate for LOOP is not statistically different from 0.
\end{table}
\noindent
We can see that the OLS estimate is biased, while the LOOP and cross estimation estimators are both unbiased.\footnote{Cross estimation is slightly biased as implemented. This is due to the difference between the out-of-bag and the leave-one-out estimates of the potential outcomes. This issue can be fixed by reducing the size of the bootstrap sample used in the random forest when making out-of-bag predictions of the potential outcomes.}  Moreover, while the true standard errors of the three methods are similar, the nominal standard errors for OLS and cross estimation are both quite biased. The nominal standard error for LOOP is also biased, but less so.  

\subsection{Simulation 2: Estimating the Treatment Effect for a Binary Response}
In our second simulation, we consider a randomized experiment in which the response is either zero or one. Each of the $N$ subjects has one of three sets of potential outcomes: (1) zero regardless of treatment assignment, (2) zero if control and one if treatment, and (3) one regardless of treatment assignment. We also have one covariate ($Z_1$) that is predictive of the outcome. Higher values of this covariate indicate that the participant is more likely to be in groups (2) or (3) than group (1). Finally, we assume there are $k$ noise covariates ($Z_k$).
\\\\
We generate $Z_1$ from a standard normal distribution. For each subject $i$, the probabilities that the subject ends up in each group is determined as follows: we calculate $w_{i1} = 1$, $w_{i2} = \exp(0.5 c \times Z_{i1})$, and $w_{i3} = \exp(c \times Z_{i1})$, where $c$ is a positive constant. The probability that observation $i$ is assigned to group $j$ is $p_{ij} = w_{ij}/(w_{i1}+w_{i2}+w_{i3})$. Thus, higher values of $c$ indicate $Z_1$ is more predictive of outcome. In addition, observation $i$ is most likely to be in the third group (and least likely to be in the first group) if $Z_{i1}$ is positive. 
\\\\
Under this framework, we consider three sets of simulations. First, we assume that both the number of subjects ($N = 200$) and the predictive power of $Z_1$ ($c=3$) are constant, and vary the number of noise covariates (from $k = 5$ to $k = 100$ in increments of 5). Next, we fix the predictive power of $Z_1$ ($c = 3$) and the number of noise covariates ($k = 50$), and vary the number of subjects from 100 to 1000 in increments of 50. Finally, we fix the number of subjects ($N=200$) and noise covariates ($k = 50$), and vary the predictive power of $Z_1$ (from $c = 1$ to $c = 5.5$ in increments of 0.5). For each simulation, we run 1,000 trials and calculate the mean nominal standard errors and true standard errors as above. For each set of simulations, we index the results to the true standard error for the simple difference estimator. We show the results of our simulations below:
\begin{figure}[H]
	\centering
	\includegraphics[width=.45\linewidth]{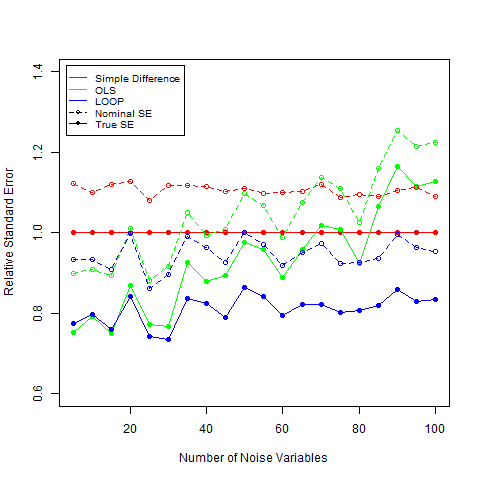}\quad
	\includegraphics[width=.45\linewidth]{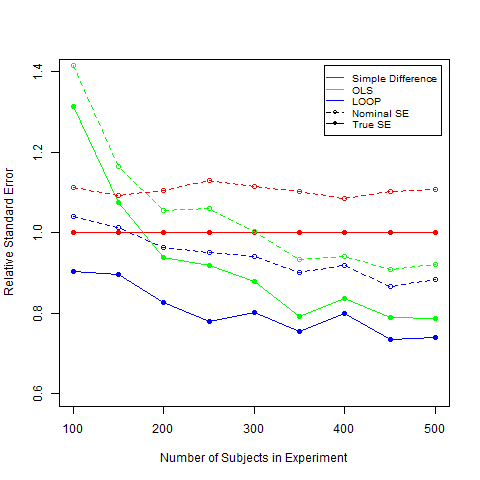}
	
	\medskip
	
	\includegraphics[width=.45\linewidth]{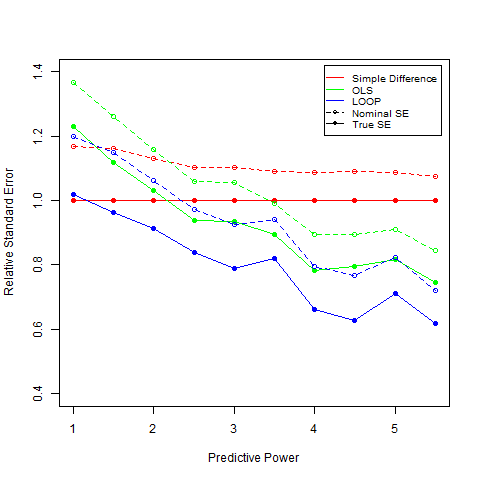}
	
	\caption{Comparison of Standard Errors for simulations}
	\label{sim2results}
\end{figure}
\noindent
We observe that while the performance of OLS declines as the number of noise covariates increases, the performance of LOOP remains constant relative to the simple difference estimator. Similarly, OLS performs worse than the simple difference estimator when the number of subjects is small, while the LOOP estimator outperforms the simple difference estimator for all sample sizes. Finally, it is important to note that covariate adjustment does not help when the covariates are not useful for predicting the outcomes. When $Z_1$ is predictive of the outcome, LOOP outperforms the simple difference estimator. However, we note that even when $Z_1$ is not predictive of outcome, the performance of the LOOP estimator is still comparable to that of the simple difference estimator. We discuss this further in the next section, where we apply the LOOP estimator to actual experimental data.
\subsection{Cash Transfer Programs and Enrollment}
In their experiment in 2005, Barrera-Osorio et al.\ studied the effects of several conditional cash transfer programs on educational outcomes for students in Bogota, Colombia. They conducted experiments in two localities of Bogota, San Cristobal and Suba. For our analysis, we focus on the San Cristobal experiment. The San Cristobal experiment involved 10,907 students from grades 6 to 11. These students were selected by lottery to be assigned to one of two treatments or to control: 3,427 students were assigned to the ``basic" treatment, 3,424 to the ``savings" treatment, and the remaining 4,056 were assigned to control. In the basic treatment, each student received a bi-monthly payment of roughly 15 USD so long as the student attended school at least 80\% of days that month. In the savings treatment, each student received a bi-monthly payment of roughly 10 USD so long as they met the attendance threshold. The remaining third was held in a bank account and paid to the students' families when it was time to re-enroll for the subsequent year. For each student, there are also various demographic covariates available.
\\\\
In their experiment, Barrera-Osorio et al.\ collected re-enrollment status from administrative records. However, they were unable to obtain re-enrollment status for approximately 10\% of the observations. In our analysis, we consider both re-enrollment status itself and whether the re-enrollment status is missing as outcome variables. For each outcome variable, we estimate the ATE for the basic treatment compared to the savings treatment, the basic treatment compared to control, and the savings treatment compared to control. We use the same covariates and restrict our analysis to students in grades 6 through 10 as in \cite{Barrera}. We provide our results in Table \ref{barreratable} below:
\begin{table}[H] \centering 
	\caption{Effect of Treatment on Missing Status and Re-enrollment Status} 
	\label{barreratable} 
	\begin{tabular}{@{\extracolsep{5pt}}llcccc} 
		\\[-1.8ex]\hline 
		\hline \\[-1.8ex] 
				 & & \multicolumn{2}{c}{Missing Status} & \multicolumn{2}{c}{Re-enrollment Status} \\ 
		Treatments & \multicolumn{1}{l}{Method} & \multicolumn{1}{c}{Estimate} & \multicolumn{1}{c}{Variance} & \multicolumn{1}{c}{Estimate} & \multicolumn{1}{c}{Variance} \\ 
		\hline \\[-1.8ex] 
		\multirow{3}{*}{Basic vs. Savings} & LOOP & -0.001057 & 0.000036 & -0.02556 & 0.00014 \\ 
		& Simple Difference & 0.006659 & 0.000055  & -0.02832 & 0.00014 \\   
		& OLS & 0.003800 & 0.000040 & -0.02941 & 0.00013 \\  \hline
		\multirow{3}{*}{Basic vs. Control} & LOOP & -0.002036 & 0.000033 & 0.01562 & 0.00013 \\ 
		& Simple Difference & 0.004128 & 0.000051 & 0.01714 & 0.00014 \\  
		& OLS & 0.001350 & 0.000037 & 0.01579 & 0.00013 \\  \hline
		\multirow{3}{*}{Saving vs. Control} & LOOP & -0.001288 & 0.000033 & 0.04219 & 0.00013 \\
		& Simple Difference & -0.002531 & 0.000049 & 0.04547 & 0.00013 \\ 
		& OLS & -0.002282 & 0.000037 & 0.04633 & 0.00013 \\  \hline
		\hline \\[-1.8ex] 
	\end{tabular} 
\end{table}
\noindent
As we see above, OLS and LOOP both provide improvement over the simple difference estimator when missing status is the outcome variable of interest. However, covariate adjustment does not help when re-enrollment status is the outcome variable, as the covariates are less predictive of outcome.

\section{Discussion} \label{discussion}
While methods of covariate adjustment can improve the precision of the estimate of the average treatment effect, they often require the researchers to perform variable selection. For example, when using post-stratification, we must be careful not to use too many covariates otherwise we partition the data set too finely. Over-adjustment can result in poorer performance with linear regression as well: OLS performs poorly when the sample size is large relative to the number of covariates or as the number of noise covariates increases.
\\\\
The LOOP estimator is an unbiased estimate of the average treatment effect and randomization justifies the assumptions made. One advantage of the LOOP estimator is that estimation of $m_i$ is very flexible. One can impute the potential outcomes using any method, so long as $\hat{m}_i$ and $T_i$ are independent. One baseline approach is to estimate $m_i$ without making use of covariates, simply taking the mean of the observed outcomes in each treatment group. In this case, the LOOP estimator is exactly equal to the simple difference estimator. This suggests that the LOOP estimator will generally outperform the simple difference estimator, so long as we use a sensible method for imputing the potential outcomes. For example, one could estimate $m_i$ using a decision tree (resulting in a post-stratified estimator) or $k$-Nearest Neighbors.
\\\\
In this paper, we suggest the use of random forests to impute the potential outcomes, as they are computationally efficient relative to other methods, improve performance over the post-stratified estimate, and allow for automatic variable selection. Because of the automatic variable selection, we can adjust for covariates without knowing ahead of time which covariates we wish to use. If the covariates are predictive of outcome, covariate adjustment with the LOOP estimator allows for improved precision over the simple difference estimator. However, even when the covariates are not predictive of outcome, the LOOP estimator generally performs as well as the simple difference estimator. Furthermore, researchers are often concerned with the validity of statistical inference after model selection. Because model selection occurs in a ``black box" with our method, any post-selection inference is still valid. In particular, when imputing the potential outcomes using random forests, the researcher will not have to do any manual variable selection and can take advantage of the automatic variable selection.

\section{Implementation in R}
The LOOP estimator is implemented in R as the \texttt{loop.estimator} package and is available on GitHub:  \texttt{https://github.com/wuje/LOOP}.

\section{Acknowledgements}
We would like to thank Yotam Shem-Tov and Luke Miratrix for helpful comments and suggestions.

\appendix
\section{Negligibility of $\bar{\gamma}$} \label{gammabar}

In this section we consider the behavior of $\bar{\gamma}$ as the sample size $N$ grows large.  In our model, the potential outcomes and the covariates are fixed parameters; they are not drawn from some probability distribution.  Thus, when we speak of a growing sample size, we must imagine a growing set of parameters.  Without any regularity conditions on these parameters, very little can be said, and thus some regularity conditions are necessary.  However, we will not propose any specific set of regularity conditions \textit{per se}, but rather we will assume that under some unspecified conditions that are appropriate for the imputation method under consideration (e.g. OLS, random forests, etc.), the following two assumptions hold: 

\begin{assumption}
There exists a constant $q$ such that for every $i$ there exists a constant $\alpha_i$ such that for all $N$, and for all $i \le N$ 
\begin{equation} 
\mathrm{Var}(\hat{m}_i) \le \frac{\alpha_i}{N^q} \label{varassumption1}
\end{equation}
(note that when $i > N$ observation $i$ is not yet in the model).
\end{assumption}
\begin{assumption}
There exist constants $C$ and $r$ such that for all $N$ 
\begin{equation} 
\max_{i \le N} \{\alpha_i\} \le CN^r. \label{varassumption2}
\end{equation}
\end{assumption}
\noindent For example, if we impute the potential outcomes using OLS, then under suitable regularity conditions assumption (\ref{varassumption1}) might hold with $q = 1$ (see \cite{Freedman}).  
Moreover, as long as the variation among the $\alpha_i$ is not too extreme, then assumption (\ref{varassumption2}) might hold for some reasonably small value of $r$.  For example, if the $\alpha_i$ follow a power law of the form
\begin{equation}
\mathrm{fraction}\{\alpha_i \ge x\} < Kx^{-\lambda} 
\end{equation}
that holds for all $N$, then assumption 2 would be satisfied with $C = K^{1/\lambda}$ and $r = 1/\lambda$; alternatively, if the tail of the distribution of the $\alpha_i$ decays exponentially, then any $r > 0$ would suffice, and if the $\alpha_i$ are bounded (which might be the case if both the response variable and the covariates are themselves bounded) then $r=0$.
In addition to the two assumptions above, we also assume in this section that $p_i = p$ for all $i$.
\\\\
Now, combining assumptions 1 and 2 results in
\begin{equation}
\mathrm{Var}(\hat{m}_i) \le \frac{C}{N^{q-r}}
\end{equation}
for all $i$, $N$ such that $i \le N$.  Together with (\ref{gammaijexpression}) from the main text, this implies that 
\begin{equation}
\gamma_{ij} \le  \frac{C\rho_{ij}}{p (1-p)N^{q-r}} 
\end{equation}
which further implies that 
\begin{equation}
\bar{\gamma} \le  \frac{C\bar{\rho}}{p (1-p)N^{q-r}} .
\end{equation}
Thus, we find that $\bar{\gamma}$ will go to 0 at a rate faster than $1/N$, allowing us to ignore the $(N-1)\bar{\gamma}$ term in (\ref{exacttauhatvar}), just as long as $\bar{\rho}$ goes to 0 at a rate faster than $1/N^{1-q+r}.$  In particular, if $q =1$, then all that is necessary is that $\bar{\rho}$ goes to zero faster than $1/N^r$; if in addition $r=0$, then all that is required is that $\bar{\rho}$ goes to zero.
\\\\
In Appendix \ref{polynomialcorrelation} we show that if $\tilde{m}_i$ is a polynomial function of degree $D$ (or smaller) for all $i$, then $\bar{\rho} \le D/(N-1)$.  Combining this fact with the arguments given above in this section, we see that for polynomial $\tilde{m}_i$, $\bar{\gamma}$ will go to zero at a rate faster than $1/N$ simply as long as 
\begin{equation}
\max_{i \le N} \mathrm{Var}(\hat{m}_i) \to 0.
\end{equation}

\section{Average correlation of $\hat{m}_iU_i$ and $\hat{m}_jU_j$ for polynomial $\hat{m}_i$} \label{polynomialcorrelation}
First, we define 
\begin{align}
\tilde{m}_i &= \hat{m}_i - \mathbb{E}(\hat{m}_i)
\end{align}
and note that 
\begin{align}
\textrm{Corr}(\hat{m}_i,\hat{m}_j) = \textrm{Corr}(\tilde{m}_i,\tilde{m}_j) = \rho_{ij}.
\end{align}
Now suppose that $\tilde{m}_i$ is a polynomial function of degree $D$ (or smaller) for all $i$.  That is, for all $i$,
\begin{equation}
\tilde{m}_i = \sum_{d=1}^D \sum_{k_1, k_2, ..., k_d} a_{i.k_1k_2...k_d}U_{k_1}U_{k_2}...U_{k_d}  \nonumber 
\end{equation}
where the second sum is over all subsets $\{k_1, k_2, ..., k_d\} \subset \{1,2,...,N\}\setminus\{i\}$.  A few comments: (1) no constant (intercept) term is needed in the expansion because $\tilde{m}_i$ has expectation 0, as do all the $U_{k_1}U_{k_2}...U_{k_d}$ terms; (2) no higher powers of the $U_k$ variables are needed, e.g. $U^3_{k}$,  since 
\begin{equation}
U^2_k = \frac{1}{p(1-p)} + \frac{1-2p}{p(1-p)}U_k \nonumber 
\end{equation}
and thus by induction any higher power of $U_k$ can be reparameterized in terms of $U_k$ itself; (3) in our notation for the coefficients $a_{i.k_1k_2...k_d}$, the ordering of the indices after the period does not matter.  In other words, there is no distinction between $a_{2.358}$, $a_{2.583}$, $a_{2.835}$, etc.  This fact will become important below when we count the number of times a specific coefficient appears in a sum.
\\\\
Note that
\begin{equation}
\mathrm{Var}(\tilde{m}_i) = \sum_{d=1}^D \sum_{k_1, k_2, ..., k_d} a^2_{i.k_1k_2...k_d}\left[\frac{1}{p(1-p)} \right]^d \nonumber 
\end{equation}
and define
\begin{equation}
b_{i.k_1k_2...k_d} = \frac{a_{i.k_1k_2...k_d}}{\sqrt{  p^d(1-p)^d   \mathrm{Var}(\tilde{m}_i)}}. \nonumber 
\end{equation}
so that
\begin{equation}
\sum_{d=1}^D \sum_{k_1, k_2, ..., k_d} b^2_{i.k_1k_2...k_d} = 1 \nonumber 
\end{equation}
and
\begin{equation}
\sum_{i=1}^N \sum_{d=1}^D \sum_{k_1, k_2, ..., k_d} b^2_{i.k_1k_2...k_d} = N  \label{bsumtoN}
\end{equation}
which is a fact we will make use of below.
\\\\
Next observe that
\begin{align}
\gamma_{ij} &= \mathrm{Cov}(\tilde{m}_i U_i, \tilde{m}_j U_j)  \nonumber \\
&= \mathbb{E} \left[ \left(\sum_{d=1}^D \sum_{k_1, k_2, ..., k_d} a_{i.k_1k_2...k_d}U_{k_1}U_{k_2}...U_{k_d}U_i \right) \left( \sum_{d=1}^D \sum_{k_1, k_2, ..., k_d} a_{j.k_1k_2...k_d}U_{k_1}U_{k_2}...U_{k_d}U_j \right) \right] \nonumber \\
&= \mathbb{E} \left[\sum_{d=1}^D \sum_{e=1}^D \sum_{k_1, k_2, ..., k_d} \sum_{l_1, l_2, ..., l_e} a_{i.k_1k_2...k_d}U_{k_1}U_{k_2}...U_{k_d} U_i a_{j.l_1l_2...l_e}U_{l_1}U_{l_2}...U_{l_e} U_j \right] \nonumber  \\
&= \sum_{d=1}^D \sum_{e=1}^D \sum_{k_1, k_2, ..., k_d} \sum_{l_1, l_2, ..., l_e} a_{i.k_1k_2...k_d} a_{j.l_1l_2...l_e}\mathbb{E} \left(U_{k_1}U_{k_2}...U_{k_d} U_i U_{l_1}U_{l_2}...U_{l_e} U_j \right) \label{aexpansion1}  
\end{align}
where again $\{k_1, k_2, ..., k_d\} \subset \{1,2,...,N\}\setminus\{i\}$ and $\{l_1, l_2, ..., l_e\} \subset \{1,2,...,N\}\setminus\{j\}$.  But
\begin{equation}
\mathbb{E} \left(U_{k_1}U_{k_2}...U_{k_d} U_i U_{l_1}U_{l_2}...U_{l_e} U_j \right) =
\begin{cases} 
      \frac{1}{p^{d+1}(1-p)^{d+1}} & \{k_1, k_2, ..., k_d, i \} = \{l_1, l_2, ..., l_e, j \} \\
      0 & \mathrm{otherwise}
   \end{cases} \nonumber 
\end{equation}
and thus we may simplify (\ref{aexpansion1}) as
\begin{align}
\gamma_{ij} = \sum_{d=1}^D \sum_{k_1, k_2, ..., k_{d-1}} a_{i.k_1k_2...k_{d-1}j} a_{j.k_1k_2...k_{d-1}i} \frac{1}{p^{d+1}(1-p)^{d+1}}  \nonumber 
\end{align}
where now the second sum is over all subsets $\{k_1, k_2, ..., k_{d-1}\} \subset \{1,2,...,N\}\setminus\{i, j\}$.
\\\\
Next observe that 

\begin{align}
\rho_{ij} &= \frac{\gamma_{ij}}{\sqrt{\mathrm{Var}(\tilde{m}_i) \mathrm{Var}(\tilde{m}_j)}\left[p(1-p) \right]^{-1}}  \nonumber \\
&= \frac{\sum_{d=1}^D \sum_{k_1, k_2, ..., k_{d-1}} a_{i.k_1k_2...k_{d-1}j} a_{j.k_1k_2...k_{d-1}i} \left[p(1-p) \right]^{-d}}{\sqrt{\mathrm{Var}(\tilde{m}_i) \mathrm{Var}(\tilde{m}_j)}}  \nonumber \\
&= \sum_{d=1}^D \sum_{k_1, k_2, ..., k_{d-1}} b_{i.k_1k_2...k_{d-1}j} b_{j.k_1k_2...k_{d-1}i}   \nonumber 
\end{align}
and therefore
\begin{align}
\sum_{i \ne j} \rho_{ij} = \sum_{i \ne j} \sum_{d=1}^D \sum_{k_1, k_2, ..., k_{d-1}} b_{i.k_1k_2...k_{d-1}j}  b_{j.k_1k_2...k_{d-1}i}. \label{precauchy}   
\end{align}
\\\\
Consider now the following sum of squared coefficients
\begin{equation}
\sum_{i \ne j} \sum_{d=1}^D \sum_{k_1, k_2, ..., k_{d-1}} b^2_{i.k_1k_2...k_{d-1}j}  \nonumber 
\end{equation}
and observe that no single coefficient shows up in the sum any more than $D$ times.  For example, the coefficient $b_{1.234}$ will show up 3 times: once when $i=1$, $j=2$, $d=3$, and $\{k_1, k_2\} = \{3, 4\}$; once when $i=1$, $j=3$, $d=3$, and $\{k_1, k_2\} = \{2, 4\}$; and once when $i=1$, $j=4$, $d=3$, and $\{k_1, k_2\} = \{2, 3\}$.  Thus
\begin{equation}
\sum_{i \ne j} \sum_{d=1}^D \sum_{k_1, k_2, ..., k_{d-1}} b^2_{i.k_1k_2...k_{d-1}j} \le D \sum_{i=1}^N \sum_{d=1}^D \sum_{k_1, k_2, ..., k_d} b^2_{i.k_1k_2...k_d}  \nonumber 
\end{equation}
where the third sum on the left hand side is over all subsets $\{k_1, k_2, ..., k_{d-1}\} \subset \{1,2,...,N\}\setminus\{i, j\}$ and the third sum on the right hand side is over all subsets $\{k_1, k_2, ..., k_d\} \subset \{1,2,...,N\}\setminus\{i\}$.  Applying (\ref{bsumtoN}) we therefore find that
\begin{equation}
\sum_{i \ne j} \sum_{d=1}^D \sum_{k_1, k_2, ..., k_{d-1}} b^2_{i.k_1k_2...k_{d-1}j} \le D N \label{bleft}
\end{equation}
and also similarly
\begin{equation}
\sum_{i \ne j} \sum_{d=1}^D \sum_{k_1, k_2, ..., k_{d-1}} b^2_{j.k_1k_2...k_{d-1}i} \le D N. \label{bright}
\end{equation}
\\\\
Given (\ref{bleft}) and (\ref{bright}) we may now apply the Cauchy-Schwarz inequality to (\ref{precauchy}) and conclude
\begin{align}
\sum_{i \ne j} \rho_{ij} \le DN  \nonumber 
\end{align}
or 
\begin{align}
\bar{\rho} &= \frac{1}{N(N-1)} \sum_{i \ne j} \rho_{ij}  \nonumber  \\
          &\le \frac{D}{N-1}
\end{align}

\section{Estimating the Covariance of  $\hat{m}_iU_i$ and $\hat{m}_jU_j$} \label{Var}
In this section, we provide an estimate for the covariance of  $\hat{m}_iU_i$ and $\hat{m}_jU_j$. First,
\begin{align}
\mathrm{Cov}(\hat{m}_iU_i,\hat{m}_jU_j) & = \mathrm{Cov}\left[[(1-p_i)\hat{t}_i+p_i\hat{c}_i]U_i,[(1-p_j)\hat{t}_j+p_j\hat{c}_j]U_j\right] \nonumber \\
& = (1-p_i)(1-p_j)\mathrm{Cov}(\hat{t}_iU_i,\hat{t}_jU_j)+(1-p_i)p_j\mathrm{Cov}(\hat{t}_iU_i,\hat{c}_jU_j) \nonumber \\
& \hspace{0.5cm} + p_i(1-p_j)\mathrm{Cov}(\hat{c}_iU_i,\hat{t}_jU_j)+p_ip_j\mathrm{Cov}(\hat{c}_iU_i,\hat{c}_jU_j).
\end{align}
Now, we let $\hat{t}_i^{+ j}$ denote the estimate of $t_i$ including the $j$-th observation, where all the treatment assignments of the other $N-2$ observations are kept as is. Similarly, we let $\hat{t}_i^{- j}$ denote the estimate of $t_i$ excluding the $j$-th observation. Then we have
\begin{align}
\mathrm{Cov}(\hat{t}_iU_i,\hat{t}_jU_j | U_{k \notin \{i,j\}}) & = \hat{t}^{+j}_i\hat{t}^{+i}_j - \hat{t}^{-j}_i\hat{t}^{+i}_j - \hat{t}^{+j}_i\hat{t}^{-i}_j + \hat{t}^{-j}_i\hat{t}^{-i}_j \nonumber \\
& = (\hat{t}^{+j}_i-\hat{t}^{-j}_i)(\hat{t}^{+i}_j-\hat{t}^{-i}_j) \nonumber \\
\mathrm{Cov}(\hat{t}_iU_i,\hat{c}_jU_j | U_{k \notin \{i,j\}}) & = \hat{t}^{+j}_i\hat{c}^{-i}_j - \hat{t}^{-j}_i\hat{c}^{-i}_j - \hat{t}^{+j}_i\hat{c}^{+i}_j + \hat{t}^{-j}_i\hat{c}^{+i}_j \nonumber \\
& = \left(\hat{t}^{+j}_i-\hat{t}^{-j}_i\right)\left(\hat{c}^{-i}_j-\hat{c}^{+i}_j\right) \nonumber \\
\mathrm{Cov}(\hat{c}_iU_i,\hat{t}_jU_j | U_{k \notin \{i,j\}}) & = \hat{c}^{-j}_i\hat{t}^{+i}_j - \hat{c}^{+j}_i\hat{t}^{+i}_j - \hat{c}^{-j}_i\hat{t}^{-i}_j + \hat{c}^{+j}_i\hat{t}^{-i}_j \nonumber \\
& = \left(\hat{c}^{-j}_i-\hat{c}^{+j}_i\right)\left(\hat{t}^{+i}_j-\hat{t}^{-i}_j\right) \nonumber \\
\mathrm{Cov}(\hat{c}_iU_i,\hat{c}_jU_j | U_{k \notin \{i,j\}}) & = \hat{c}^{-j}_i\hat{c}^{-i}_j - \hat{c}^{+j}_i\hat{c}^{-i}_j - \hat{c}^{-j}_i\hat{c}^{+i}_j + \hat{c}^{+j}_i\hat{c}^{+i}_j \nonumber \\
& = \left(\hat{c}^{-j}_i-\hat{c}^{+j}_i\right)\left(\hat{c}^{-i}_j-\hat{c}^{+i}_j\right).
\end{align}
Note that $\hat{t}^{+j}_i$ is calculable when $T_j = 1$, but not when $T_j = 0$, as $t_j$ is not observable when $T_j = 0$. Similarly, $\hat{c}^{+j}_i$ is calculable when $T_j = 0$, but not when $T_j = 1$. Thus, we use following estimate of the covariance (where all the terms are estimable):
\begin{align}
\widehat{\mathrm{Cov}}(\hat{m}_iU_i,\hat{m}_jU_j) & =\left\{\begin{array}{lr} \frac{(1-p_i)(1-p_j)}{p_ip_j}
(\hat{t}^{+j}_i-\hat{t}^{-j}_i)(\hat{t}^{+i}_j-\hat{t}^{-i}_j), & T_i = T_j = 1 \\
(\hat{t}^{+j}_i-\hat{t}^{-j}_i)(\hat{c}^{-i}_j-\hat{c}^{+i}_j), & T_i = 0, T_j = 1 \\
(\hat{c}^{-j}_i-\hat{c}^{+j}_i)(\hat{t}^{+i}_j-\hat{t}^{-i}_j), & T_i = 1, T_j = 0 \\
\frac{p_ip_j}{(1-p_i)(1-p_j)} (\hat{c}^{-j}_i-\hat{c}^{+j}_i)(\hat{c}^{-i}_j-\hat{c}^{+i}_j), & T_i = T_j = 0
\end{array}\right.
\end{align}
which is an unbiased estimate of the covariance:
\begin{align}
& \mathbb{E}[\widehat{\mathrm{Cov}}(\hat{m}_iU_i,\hat{m}_jU_j)|U_{k \notin \{i,j\}}]  \nonumber \\
& = p_ip_j\frac{(1-p_i)(1-p_j)}{p_ip_j}
(\hat{t}^{+j}_i-\hat{t}^{-j}_i)(\hat{t}^{+i}_j-\hat{t}^{-i}_j) + (1-p_i)p_j(\hat{t}^{+j}_i-\hat{t}^{-j}_i)(\hat{c}^{-i}_j-\hat{c}^{+i}_j) \nonumber \\
& \hspace{.5cm} + p_i(1-p_j)(\hat{c}^{-j}_i-\hat{c}^{+j}_i)(\hat{t}^{+i}_j-\hat{t}^{-i}_j) + (1-p_i)(1-p_j)\frac{p_ip_j}{(1-p_i)(1-p_j)} (\hat{c}^{-j}_i-\hat{c}^{+j}_i)(\hat{c}^{-i}_j-\hat{c}^{+i}_j) \nonumber \\
& = (1-p_i)(1-p_j)
\mathrm{Cov}(\hat{t}_iU_i,\hat{t}_jU_j | U_{k \notin \{i,j\}}) + (1-p_i)p_j\mathrm{Cov}(\hat{t}_iU_i,\hat{c}_jU_j | U_{k \notin \{i,j\}}) \nonumber \\
& \hspace{.5cm} + p_i(1-p_j)\mathrm{Cov}(\hat{c}_iU_i,\hat{t}_jU_j | U_{k \notin \{i,j\}}) + p_ip_j \mathrm{Cov}(\hat{c}_iU_i,\hat{c}_jU_j | U_{k \notin \{i,j\}}) \nonumber \\
& = \mathrm{Cov}(\hat{m}_iU_i,\hat{m}_jU_j | U_{k \notin \{i,j\}}).
\end{align}
We take the expectation across all randomizations to show $\widehat{\mathrm{Cov}}(\hat{m}_iU_i,\hat{m}_jU_j)$ is unbiased.
\begin{align}
\mathbb{E}[\mathrm{Cov}(\hat{m}_iU_i,\hat{m}_jU_j | U_{k \notin \{i,j\}})]
& = \mathbb{E}[\mathbb{E}(\hat{m}_iU_i\hat{m}_jU_j | U_{k \notin \{i,j\}}) - \mathbb{E}(\hat{m}_iU_i| U_{k \notin \{i,j\}})\mathbb{E}(\hat{m}_jU_j | U_{k \notin \{i,j\}})] \nonumber \\
& = \mathbb{E}[\mathbb{E}(\hat{m}_iU_i\hat{m}_jU_j | U_{k \notin \{i,j\}})] \nonumber \\
& = \mathbb{E}(\hat{m}_iU_i\hat{m}_jU_j)] \nonumber \\
& = \mathrm{Cov}(\hat{m}_iU_i,\hat{m}_jU_j)
\end{align}
Summing across all $i, j$ pairs yields an unbiased estimate of $\sum_{i \neq j}\mathrm{Cov}(\hat{m}_iU_i,\hat{m}_jU_j)$.
\section{The Mean Squared Error of $\hat{m}_i$} \label{bound}
Below, we express MSE($\hat{m}_i$) in terms of the MSEs of $\hat{t}_i$ and $\hat{c}_i$:
\begin{align} 
\mathrm{MSE}(\hat{m}_i) & = [\mathbb{E}(\hat{m}_i - m_i)]^2 + \mathrm{Var}(\hat{m}_i) \nonumber
\\
& = \left[\mathbb{E}[(1-p)\hat{t}_i + p\hat{c}_i - (1-p)t_i - pc_i]\right]^2 + \mathrm{Var}[(1-p)\hat{t_i} + p\hat{c_i}] \nonumber
\\ & = [\mathbb{E}[(1-p)(\hat{t}_i - t_i)] + \mathbb{E}[p(\hat{c}_i -c_i)]]^2 + (1-p)^2\mathrm{Var}(\hat{t_i})+p^2\mathrm{Var}(\hat{c_i}) \nonumber \\
& \hspace{.75cm}+2p(1-p)\mathrm{Cov}(\hat{t_i},\hat{c_i}) \nonumber
\\ & = [(1-p)\mathrm{Bias}(\hat{t}_i) + p\mathrm{Bias}(\hat{c}_i)]^2 + (1-p)^2\mathrm{Var}(\hat{t_i})+p^2\mathrm{Var}(\hat{c_i})+2p(1-p)\mathrm{Cov}(\hat{t_i},\hat{c_i}) \nonumber
\\ & = (1-p)^2\mathrm{Bias}^2(\hat{t}_i) + p^2\mathrm{Bias}^2(\hat{c}_i) + 2p(1-p) \mathrm{Bias}(\hat{t}_i)\mathrm{Bias}(\hat{c}_i)  \nonumber \\
& \hspace{.75cm} + (1-p)^2\mathrm{Var}(\hat{t_i})+p^2\mathrm{Var}(\hat{c_i})+2p(1-p)\mathrm{Cov}(\hat{t_i},\hat{c_i})] \nonumber
\\ & = (1-p)^2\mathrm{MSE}(\hat{t_i})+p^2\mathrm{MSE}(\hat{c_i})+2p(1-p)\left[\mathrm{Cov}(\hat{t_i},\hat{c_i})+ \mathrm{Bias}(\hat{t}_i)\mathrm{Bias}(\hat{c}_i)\right] \nonumber
\\ &
\leq (1-p)^2\mathrm{MSE}(\hat{t_i})+p^2\mathrm{MSE}(\hat{c_i})+2p(1-p)\sqrt{\mathrm{MSE}(\hat{t_i})\mathrm{MSE}(\hat{c_i})}. \label{c_inequal}
\end{align}
To show inequality (\ref{c_inequal}), we prove that:
\begin{align}
\mathrm{Cov}(\hat{t_i},\hat{c_i})+ \mathrm{Bias}(\hat{t}_i)\mathrm{Bias}(\hat{c}_i) \leq \sqrt{\mathrm{MSE}(\hat{t_i})\mathrm{MSE}(\hat{c_i})}.
\end{align}
The proof is trivial, but is included here for the sake of completeness.
\begin{proof}
	Let $\mathrm{Cov}(\hat{t_i},\hat{c_i}) = C$, $\mathrm{Bias}(\hat{t}_i) = B_t$, $\mathrm{Bias}(\hat{c}_i) = B_c$, $\mathrm{Var}(\hat{t}_i) = V_t$, $\mathrm{Var}(\hat{c}_i) = V_c$:
	\begin{align} 
	C+B_tB_c & \leq\sqrt{\mathrm{MSE}(\hat{t_i})\mathrm{MSE}(\hat{c_i})} \nonumber 
	\\
	(C+B_tB_c)^2 & \leq (B_t^2+V_t)(B_c^2+V_c) \nonumber 
	\\
	C^2 + 2CB_tB_c + B_t^2B_c^2 & \leq V_tV_c + V_tB_c^2 + V_cB_t^2 +B_t^2B_c^2 \nonumber 
	\\
	C^2 + 2CB_tB_c & \leq V_tV_c + V_tB_c^2 + V_cB_t^2.
	\end{align}
	$\mathrm{Cov}(\hat{t_i},\hat{c_i})$ is less than or equal to $\sqrt{\mathrm{Var}(\hat{t_i})\mathrm{Var}(\hat{c_i})}$ so it is sufficient to show:
	\begin{align} 
	V_tV_c + 2\sqrt{V_tV_c}B_tB_c & \leq V_tV_c + V_tB_c^2 + V_cB_t^2
	\nonumber \\
	2\sqrt{V_tV_c}B_tB_c & \leq V_tB_c^2 + V_cB_t^2
	\nonumber \\
	0 & \leq V_tB_c^2 - 2\sqrt{V_tV_c}B_tB_c + V_cB_t^2
	\nonumber \\
	0 & \leq (\sqrt{V_t}B_c - \sqrt{V_c}B_t)^2.
	\end{align}
\end{proof}

\section{$\hat{M}_t$ and $\hat{M}_c$ are Approximately Unbiased} \label{MSEestimation}
Recall that 
\begin{equation}
M_t = \frac{1}{N}\sum_{i=1}^{N}\mathrm{MSE}(\hat{t_i})
\end{equation}
and
\begin{equation}
M_c = \frac{1}{N}\sum_{i=1}^{N}\mathrm{MSE}(\hat{c_i}).
\end{equation}
and that 
\begin{equation} 
\hat{M}_t = \frac{1}{n}\sum_{i \in \mathcal{T}}(\hat{t}_i - t_i)^2
\end{equation}
and
\begin{equation}
\hat{M}_c = \frac{1}{N-n}\sum_{i \in \mathcal{C}}(\hat{c}_i - c_i)^2. 
\end{equation}
Consider also the estimators
\begin{equation} 
\tilde{M}_t = \frac{1}{Np}\sum_{i \in \mathcal{T}}(\hat{t}_i - t_i)^2
\end{equation}
and
\begin{equation}
\tilde{M}_c = \frac{1}{N(1-p)}\sum_{i \in \mathcal{C}}(\hat{c}_i - c_i)^2. 
\end{equation}
We will show that $\tilde{M}_t$ and $\tilde{M}_c$ are exactly unbiased.  From this it follows that in large samples, $\hat{M}_t$ and $\hat{M}_c$ are nearly unbiased, since with high probability $n \approx Np$ and $N-n \approx N(1-p)$.  

\begin{align}
\mathbb{E}(\tilde{M}_t) &= \mathbb{E}\left[\frac{1}{Np}\sum_{i \in \mathcal{T}}(\hat{t}_i - t_i)^2\right] \nonumber \\
&=  \mathbb{E}\left[\frac{1}{Np}  \sum_{i =1}^NT_i(\hat{t}_i - t_i)^2\right] \nonumber \\
&=  \frac{1}{Np}  \sum_{i =1}^N \mathbb{E}\left[ T_i(\hat{t}_i - t_i)^2\right] \nonumber \\
&=  \frac{1}{Np}  \sum_{i =1}^N \mathbb{E}\left( T_i \right) \mathbb{E}\left[(\hat{t}_i - t_i)^2\right] \nonumber \\
&=  \frac{1}{Np}  \sum_{i =1}^N p \mathbb{E}\left[(\hat{t}_i - t_i)^2\right] \nonumber \\
&=  \frac{1}{N}  \sum_{i =1}^N \mathrm{MSE}\left(\hat{t}_i\right)
\end{align}
The argument for $\tilde{M}_c$ is analogous.

\section{The Relationship between $\hat{M}_t$ and the Sample Variance of $\hat{t}_i$} \label{M_t Sample Var}
We show that $\hat{M}_t = \frac{n}{n-1}\frac{1}{n-1}\sum_{i \in \mathcal{T}} (t_i-\bar{t})^2$. Without loss of generality, assume that $\mathcal{T} = \{1, ..., n\}$:

\begin{align}
\hat{M}_t 
& = \frac{1}{n}\sum_{i = 1}^{n} (\hat{t}_i - t_i)^2 \nonumber \\
& = \frac{1}{n}\sum_{i = 1}^{n} (\hat{t}_i^2 - 2\hat{t}_it_i + t_i^2) \label{53}
\end{align}
We deal with the first two terms:

\begin{align}
\sum_{i = 1}^{n} \hat{t}_i^2 - 2\sum_{i = 1}^{n}\hat{t}_it_i 
& = \sum_{i = 1}^{n}\frac{(\sum_{k \neq i} t_k)^2}{(n-1)^2} - 2\sum_{i = 1}^{n}\frac{\sum_{k \neq i} t_k}{n-1}t_i \nonumber \\ 
& = \sum_{i = 1}^{n}\frac{\sum_{j,k \neq i} t_jt_k}{(n-1)^2} - 2\sum_{j \neq k}\frac{t_jt_k}{n-1} \nonumber \\ 
& = \frac{1}{(n-1)^2}\left[(n-1)\sum_{i=1}^{n}t_i^2 + (n-2)\sum_{j \neq k}t_jt_k\right] - 2(n-1)\sum_{j \neq k}\frac{t_jt_k}{(n-1)^2} \nonumber \\ 
& = \frac{\sum_{i=1}^{n}t_i^2}{n-1} + (n-2)\sum_{j \neq k}\frac{t_jt_k}{(n-1)^2} - 2(n-1)\sum_{j \neq k}\frac{t_jt_k}{(n-1)^2} \nonumber \\
& = \frac{\sum_{i=1}^{n}t_i^2}{n-1} - n\sum_{j \neq k}\frac{t_jt_k}{(n-1)^2} \label{54}
\end{align}
Plugging (\ref{54}) into (\ref{53}), we can express $\hat{M}_t$ as follows:
\begin{align}
\hat{M}_t 
& = \frac{1}{n}\left[\sum_{i = 1}^{n} \frac{t_i^2}{n-1} - n\sum_{j \neq k}\frac{t_jt_k}{(n-1)^2} + \frac{n-1}{n-1}\sum_{i = 1}^{n}t_i^2 \right] \nonumber \\
& = \frac{1}{n}\left[\sum_{i = 1}^{n} \frac{nt_i^2}{n-1} - n\sum_{j \neq k}\frac{t_jt_k}{(n-1)^2} \right] \nonumber \\
& = \frac{1}{n-1}\left[\sum_{i = 1}^{n} t_i^2 - \sum_{j \neq k}\frac{t_jt_k}{n-1} \right] \nonumber \\
& = \frac{1}{n-1}\left[\frac{n-1}{n-1}\sum_{i = 1}^{n} t_i^2 - \sum_{j \neq k}\frac{t_jt_k}{n-1} \right] \nonumber \\
& = \frac{1}{n-1}\left[\frac{n}{n-1}\sum_{i = 1}^{n} t_i^2 - \frac{\sum_{i = 1}^{n}t_i^2 + \sum_{j \neq k}t_jt_k}{n-1}\frac{n}{n} \right] \nonumber \\
& = \frac{1}{n-1}\frac{n}{n-1}\left[\sum_{i = 1}^{n} t_i^2 - \frac{(\sum_{i = 1}^{n}t_i)^2}{n} \right] \nonumber \\
& = \frac{n}{n-1}\frac{1}{n-1}\left[\sum_{i = 1}^{n} t_i^2 - n\bar{t}^2 \right]
\end{align}

\section{Random Drop Procedure Example} \label{RandomDropEx}
Consider an experiment with five participants, in which two participants are to be randomly assigned to treatment and the remaining three to control, and suppose we wish to estimate $m_1$ using the random drop procedure.  If $T_1 = 1$, we randomly pick a control observation and omit it when calculating $\hat{m}_1$. Similarly, if $T_1 = 0$ we randomly drop a treatment observation. 
\\\\
On the left side of the table below, we show the 10 possible (and equally likely) treatment assignment vectors.  The right side of the table shows the possible treatment assignment vectors after applying the random drop procedure; a backslash represents the dropped observation. For example, when the treatment assignment is 5) CTTCC, we could randomly drop either of the two treatment observations, resulting in either C$\backslash$TCC or CT$\backslash$CC.
\begin{table}[H]
	\begin{center}
		\begin{tabular}{l | c | c c c }
			\hline
			\# & Treatment Assignments & Potential Drops & & \\
			\hline
			1) & T T C C C & T T $\backslash$ C C & T T C $\backslash$ C & T T C C $\backslash$ \\
			2) & T C T C C & T $\backslash$ T C C & T C T $\backslash$ C & T C T C $\backslash$ \\
			3) & T C C T C & T $\backslash$ C T C & T C $\backslash$ T C & T C C T $\backslash$ \\
			4) & T C C C T & T $\backslash$ C C T & T C $\backslash$ C T & T C C $\backslash$ T \\
			5) & C T T C C & C $\backslash$ T C C & C T $\backslash$ C C & \\
			6) & C T C T C & C $\backslash$ C T C & C T C $\backslash$ C & \\
			7) & C T C C T & C $\backslash$ C C T & C T C C $\backslash$ & \\
			8) & C C T T C & C C $\backslash$ T C & C C T $\backslash$ C & \\
			9) & C C T C T & C C $\backslash$ C T & C C T C $\backslash$ & \\
			10) & C C C T T & C C C $\backslash$ T & C C C T $\backslash$ & \\
		\end{tabular}
		\label{randomdroptbl}
	\end{center}
\end{table}
\noindent We can use the above example to illustrate how $\hat{m}_1$ is independent of $T_1$. Regardless of whether $T_1$ is 0 or 1, we calculate $\hat{m}_1$ using a single treatment observation and two control observations; moreover, the value of $T_1$ does not tell us anything about which two of the four possible units will be in control, or which one of the four will be in treatment.   
\\\\
For example, consider the arrangement T$\backslash$CC for the last four observations. We can see that this arrangement occurs in exactly one in twelve of the combinations where $T_1 = 1$ and one in twelve of the combinations where $T_1 = 0$. That is, 
$$\mathrm{P}( T\backslash CC | T_1 = 1) = \mathrm{P}( T\backslash CC | T_1 = 0) = 1/12.$$
The same is true of all of the other 11 possible arrangements of the last four observations.  Thus $T_1$ and $\hat{m}_1$ are independent.

\section{Expectation of the Random Drop Procedure} \label{RandomDrop}
In this section, we show that $\hat{\tau}$ remains relatively unchanged by the random drop procedure in the case where we estimate $m_i$ without using covariates. To do this, we show that the expectation (over random drops) of the estimate of the average treatment effect obtained from the random drop procedure is exactly equal to the estimate had we not used the random drop procedure at all.
\\\\
Consider the case where we estimate $m_i$ without using covariates. That is, we impute $t_i$ as the average of the treated units and $c_i$ as the average of the control units (omitting observation $i$ each time). If unit $i$ was in the control group, then each time we estimate $m_i$, we would drop a random observation in the treatment group before taking the averages of the observed outcomes. While we could repeat this procedure many times and average the resulting estimates to get our final estimate of $m_i$, we could instead take the expected value of the ``random drop" estimate over all possible drops. In this case, the value of $\hat{m}_i$ is exactly equal to the estimate had we not dropped any observations in the first place. Without loss of generality, we assume that observation $i$ is assigned to control. Let $\hat{m}_{i,-k}$ and $\hat{\tau}_{i,-k}$ denote the estimates where we randomly dropped the $k$-th observation and let $\hat{m}_{i,\cdot}$ and $\hat{\tau}_{i,\cdot}$ denote their expected values over all possible drops.
\begin{align}
\mathbb{E}_k(\hat{m}_{i,-k}) & = \frac{1}{n}\sum_{k \in \mathcal{T}} \hat{m}_{i,-k} \nonumber \\
& = \frac{1}{n}\sum_{k \in \mathcal{T}} \left[\frac{\sum_{j \in \mathcal{T} \backslash \{i,k\}} Y_j}{n-1} + \frac{\sum_{j \in \mathcal{C} \backslash \{i,k\}} Y_j}{N-n-1} \right] \nonumber \\
& = \frac{1}{n}\sum_{k \in \mathcal{T}} \left[\frac{\sum_{j \in \mathcal{T} \backslash \{k\}} Y_j}{n-1}\right] + \frac{1}{n}\sum_{k \in \mathcal{T}} \left[\frac{\sum_{j \in \mathcal{C} \backslash \{i\}} Y_j}{N-n-1} \right] \nonumber \\ 
& = \frac{1}{n} \left[\frac{(n-1)\sum_{j \in \mathcal{T}} Y_j}{n-1}\right] + \frac{1}{n}\left[\frac{n\sum_{j \in \mathcal{C} \backslash \{i\}} Y_j}{N-n-1} \right] \nonumber \\ 
& = \frac{\sum_{j \in \mathcal{T}} Y_j}{n} + \frac{\sum_{j \in \mathcal{C} \backslash \{i\}} Y_j}{N-n-1}.
\end{align}
This last line is equal to the value of $\hat{m}_i$ that we would have gotten had we not dropped any observations besides $i$.  Our estimate for $\hat{\tau}$ would also be the same as if we had not used the random drop procedure (\textit{i.e.}, $\mathbb{E}_k(\hat{\tau}_{i,-k}) = \hat{\tau}_i$). A similar argument can be used to show that if we were to use the random drop procedure when estimating $\hat{m}_i$ using a decision tree, the expected value of $\hat{\tau}$ would still be the post-stratified estimate.

\end{document}